\documentclass[11pt]{article}
\usepackage{moriond}

\def\be{\begin{equation}}
\def\ee{\end{equation}}
\def\bea{\begin{eqnarray}}
\def\eea{\end{eqnarray}}

\newcommand{\Photo}{}

\begin{document}
\vspace*{4cm}
\title{GRAVITATIONAL LENS SYSTEM B0218+357: CONSTRAINTS ON LENS MODEL AND HUBBLE CONSTANT}

\author{ T.I. LARCHENKOVA (1), A.A. LUTOVINOV (2), N.S. LYSKOVA (2)}

\address{(1) ASC of P.N.Lebedev Physical Institute, Leninskiy prospect 53, Moscow, Russia\\
(2) Space Research Institute, Profsoyuznaya 84/32, Moscow, Russia}

\maketitle\abstracts{
The possibility to use observations of lensed relativistic jets on very small angular scales to
construct proper models of spiral lens galaxies and to independently determine the Hubble
constant is considered. On the example of the system B0218+357 it is shown that there exists a
great choice of model parameters adequately reproducing its observed large-scale
properties but leading to a significant spread in the Hubble constant. The jet image
position angle is suggested as an additional parameter that allows the range of models
under consideration to be limited. It is shown that the models for which the jet image
position angles differ by at least 40$^\circ$ can be distinguished between themselves during
observations on very small angular scales. Observations with the space interferometer
RadioAstron-Earth is proposed to determine the jet orientation.
}

Gravitational lens observations make it possible to estimate the most important cosmological
parameter, the Hubble constant, if the surface density distribution of the lens, time delay
between source images, the relative position of the lens and the source are known. The source
B0218+357 ($z = 0.96$~\cite{Patnaik}) with a large-scale jet is a ''gold lens'' for the aforesaid goal
because the time delay between the two compact core images has been measured with a good
accuracy~\cite{Biggs},\cite{Cohen} and there is no gravitational field distortion
from nearby sources. But at the same time there is a significant uncertainty in determining the relative position
of the lens (spiral galaxy) and the source~\cite{York}, which cannot be measure with present-day
optic and infrared instruments due to a small angular separation of them. Therefore it is
necessary to construct an accurate model of the system to evaluate the Hubble constant. It was
shown that there exists a fairly wide choice of model parameters adequately reproducing the large
scale observed properties of B0218+357 (the intensity ratio of images A and B for the compact
source $I_A/I_B\simeq(3.1-3.7)$, the image separation $d \simeq 335$ milliarcseconds (mas), the
position angle of the large-scale jet) but leading to a significant spread in the Hubble
constant~\cite{Larchenkova_b}. Furthermore the ring-like structure (see Fig.1, left panel~\cite{Biggs}) observed in
the radio band and produced by the lensing of the large-scale jet appears only for a limited set
of model parameters and jet directions (see, e.g., Larchenkova et al.~\cite{Larchenkova_a}).
However even for this limited set of parameter the spread in
the estimates of the Hubble constant is still significant. Here we propose to observe the jet image position angle on
scales of tens of microarcseconds ($\mu$as) and use it as an additional parameter to restrict
models remaining after the large-scale analysis. The main idea of the proposition is that the
lensed jet images of source B0218+357 retain its proper geometric shape on scales of
tens of $\mu$as.

According to the VLBA survey of extragalactic sources at 15\,GHz the correlated flux density for
B0218+357 does not decrease to zero even at the maximum projected baseline of this ground-based
interferometer ($440\times10^{6}$ wavelengths) but is about 250 mJy~\cite{Kovalev}. At 5\,GHz the
components may be more extended, but we know that they are still visible well at
8.4\,GHz~\cite{Biggs}. Thus, observations of this source with a long baseline space interferometer
are reasoned from a viewpoint of determining the structure and the direction of the relativistic
jet at very small angular distances from the nozzle (the ejector of relativistic particles).

Fig.1 (right panel) shows the visibility function for four sets of model parameters with
different position angles (Fig.1, middle panel) of the jet emerging from the brighter
compact core image (image A) in January 2014 (see details in Larchenkova et
al.~\cite{Larchenkova_b}). The gray area corresponds to the projected baselines of the space
VLBI Mission RadioAstron (RA) with Effelsberg, Arecibo and Evptoria radiotelescopes at 5 GHz. From this figure it is
clearly seen, that the possibility to distinguish one model from the other will depend on the
signal-to-noise ratio. Taking into account the current RA detection threshold ($\sim 7\sigma$),
the models for which the jet image position angles differ by at least 40$^\circ$ can be
distinguished between themselves. It is necessary to note that a probable non-uniformity of the
intensity distribution along the jet (e.g., a presence of knots) practically have not an effect on this conclusion.

\begin{figure}
\hbox{
\includegraphics[width=0.23\textwidth,bb=227 279 380 535,clip]{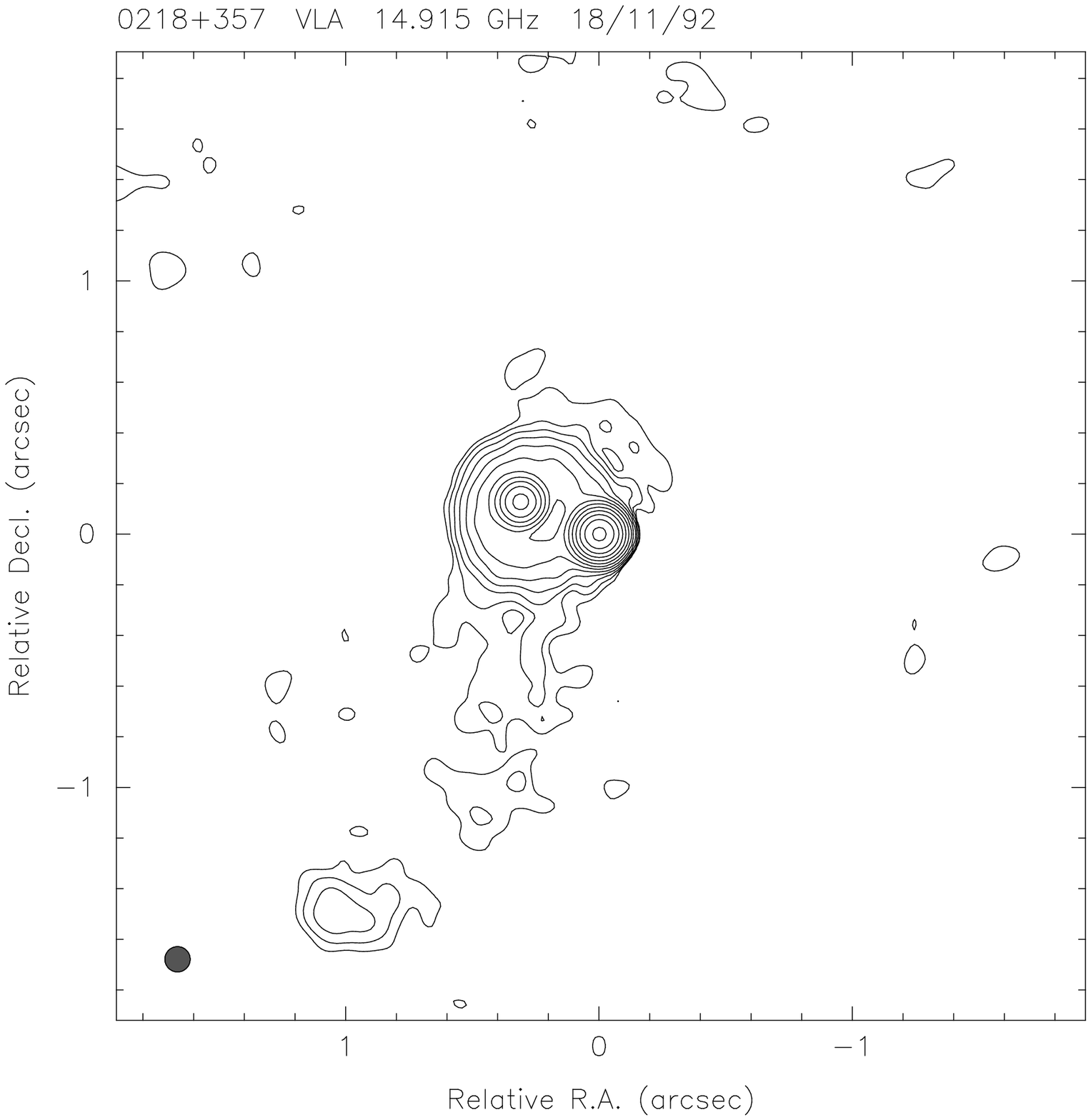}
\includegraphics[width=0.38\textwidth,bb=50 355 510 800,clip]{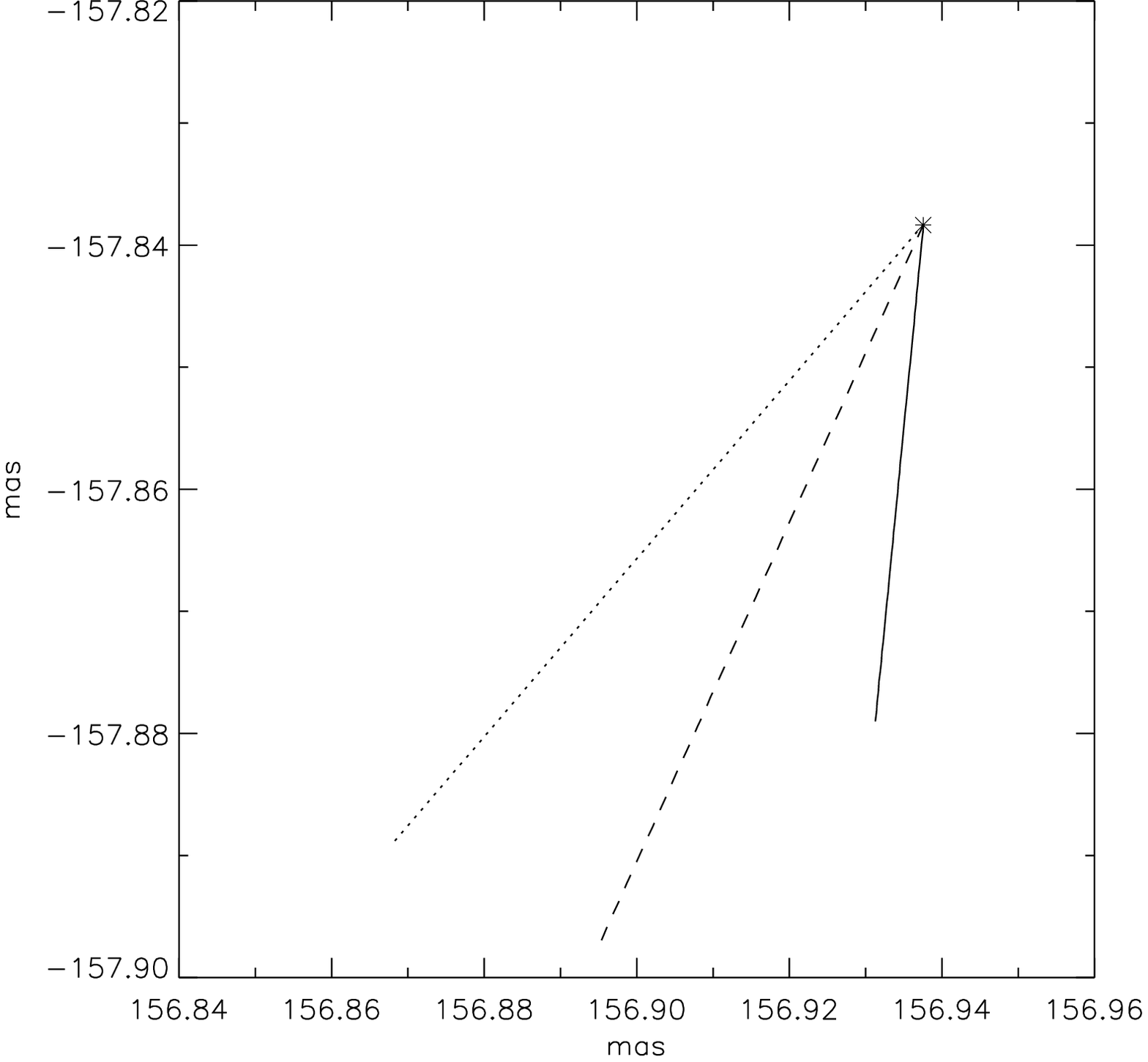}
\includegraphics[width=0.36\textwidth,bb=535 365 955 810,clip]{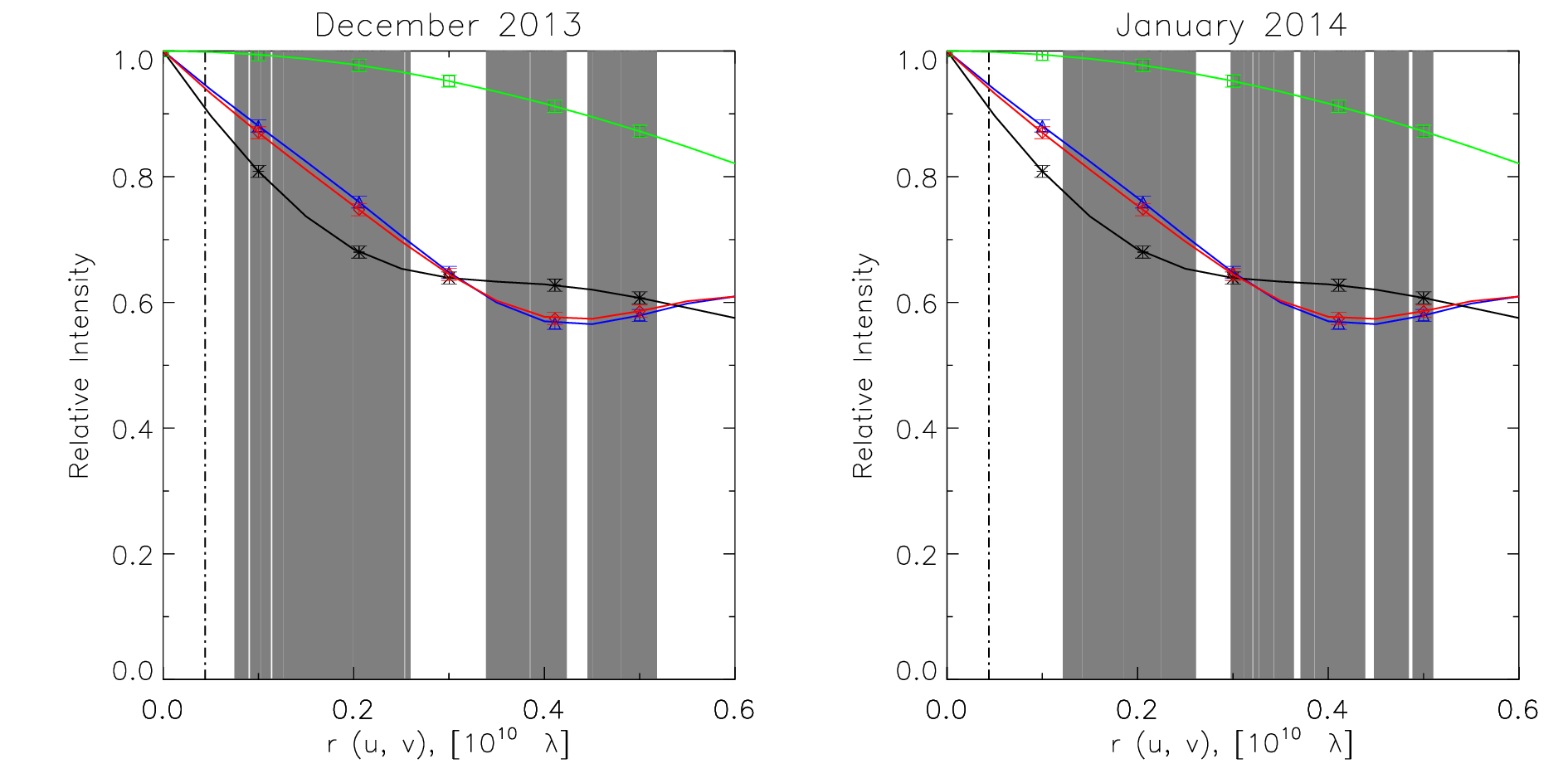}
}
\caption{\small Left: VLA 15 GHz radio map of B0218+357. Clearly visible
are the two compact cores (A to the right), the Einstein ring and non-lensed large-scale radio
jet to the south. Middle: The initial phase (30 $\mu$as) of the relativistic jet emerging from
image A for the three different position angles. Right: The visibility function for the four sets of model parameters
with position angles corresponding middle panel with projected baselines of the space VLBI Mission
RadioAstron with the Effelsberg, Arecibo and Evpatoria radiotelescopes at 5\,GHz (gray areas) in
January 2014. The typical uncertainties of measurements with RadioAstron are shown by vertical lines.}
\end{figure}

\section*{Acknowledgments}

The study was supported by The Ministry of education and science of Russian Federation, projects N8405,
N8701 and grant of NSh-2915.2012.2.


\section*{References}

\begin{thebibliography}{99}
\bibitem{Patnaik} A. Patnaik {\it et al}, {\em NNRAS} {\bf 274}, L5 (1995)

\bibitem{Biggs} A. Biggs {\it et al}, {\em NNRAS} {\bf 304}, 349 (1999)

\bibitem{Cohen} A. Cohen {\it et al}, {\em ApJ} {\bf 545}, 578 (2000)

\bibitem{York} T. York {\it et al}, {\em MNRAS} {\bf 357}, 124 (2005)

\bibitem{Larchenkova_b} T. Larchenkova {\it et al}, {\em Astron. Lett.} {\bf 37}, 441 (2011a)

\bibitem{Larchenkova_a} T. Larchenkova {\it et al}, {\em Astron. Lett.} {\bf 37}, 233 (2011b)

\bibitem{Kovalev} Y. Kovalev {\it et al}, {\em AJ} {\bf 130}, 2473 (2005)

\end{thebibliography}
\end{document}